\documentclass[12pt]{revtex4-2}

\usepackage{natbib}
\usepackage{amsmath}
\usepackage{xcolor}
\usepackage{tikz}
\usepackage{comment}
\usepackage[absolute,overlay]{textpos}

\usepackage{graphicx}
\usepackage{dcolumn}
\usepackage{bm}

\begin{document}

\begin{textblock*}{15cm}(10cm,26cm) 
   \textcolor{red}{Published in Physical Review E \textbf{109},  014203 (2024)}
\end{textblock*}

\preprint{APS/123-QED}

\title{Hamming distance as a measure of spatial chaos in evolutionary games}

\author{Gaspar Alfaro}
 \email{gaspar.alfaro@urjc.es}
\affiliation{Nonlinear Dynamics, Chaos and Complex Systems Group, Departamento de  Física, Universidad Rey Juan Carlos, Tulipán s/n, Móstoles, 28933, Madrid, Spain}
\author{Miguel A.F. Sanjuán}
 \email{miguel.sanjuan@urjc.es}
\affiliation{Nonlinear Dynamics, Chaos and Complex Systems Group, Departamento de  Física, Universidad Rey Juan Carlos, Tulipán s/n, Móstoles, 28933, Madrid, Spain}

\date{\today}

\begin{abstract}

From a context of evolutionary dynamics, social games can be studied as complex systems that may converge to a Nash equilibrium. Nonetheless, they can behave in an unpredictable manner when looking at the spatial patterns formed by the agents' strategies. This is known in the literature as spatial chaos. In this paper we analyze the problem for a deterministic prisoner's dilemma and a public goods game and calculate the Hamming distance that separates two solutions that start at very similar initial conditions for both cases. The rapid growth of this distance indicates the high sensitivity to initial conditions, which is a well-known indicator of chaotic dynamics.

\end{abstract}

\keywords {
evolutionary dynamics \sep social games \sep prisoner's dilemma \sep public goods  \sep numerical simulations \sep Hamming distance 
 }

\maketitle

\begin{tikzpicture}[]
(first)
    node {Blue circle behind text};
\end{tikzpicture}

\section{Introduction}

Evolutionary Game Theory has granted us a better knowledge of social behavior, for example understanding of cooperation, which has a great importance in social games like the prisoner's dilemma \cite{Prisionero1,Prisionero2}, PD, or the public goods game \cite{PublicGoods1,PublicGoods2}, PGG. We think of an evolutionary game as a dynamical problem of natural selection, where the payoff of each agent, acting as a fitness, depends on its strategy and on that of its neighbors. This field has attracted physicists \cite{SocialPhy} that have included ideas from nonlinear dynamics, like the discovery of spatial, pattern-forming chaos arising from a simple deterministic game as in the research work of Martin A. Nowak and Robert M. May \cite{SpatialChaos}.

The rules of the game are simple: there are cooperators, C,  and defectors, D, which play games with one neighbor. They gain a payoff according to the payoff matrix:

\begin{table}[h]
    \begin{tabular}{c|c c}
        ~& ~C~ & ~D~  \\
        \hline
        ~C~ & R & S  \\
        ~D~ & T & P
    \end{tabular}
    \label{PayoffMatrix}
\end{table}


The game is a prisoner's dilemma when $T>R>P>S$. This is called a dilemma because even though cooperation benefits all players in the group, defection is the preferable strategy. This is known as tragedy of the commons \cite{TragedyCommons}. But all is not lost for cooperation, since it can prosper under several conditions like reputation effects \cite{ReputationEffects1,ReputationEffects2} and network reciprocity \cite{NetworkReciprocity1,NetworkReciprocity2,NetworkReciprocity3}. 

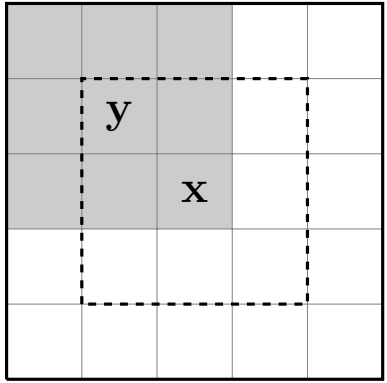
\begin{figure}
\centering
\begin{tikzpicture}

\fill[gray!40!white] (0,2) rectangle (3,5);

\draw[step=1cm,gray,very thin] (0,0) grid (5,5);

\draw[very thick] (0,0) -- (0,5) -- (5,5) -- (5,0) -- (0,0);
\draw[dashed, very thick] (1,1) -- (1,4) -- (4,4) -- (4,1) -- (1,1);

\draw (2.5,2.5) node {\Large{\textbf{x}}};
\draw (1.5,3.5) node {\Large{\textbf{y}}};

\end{tikzpicture}
\caption{Agent $x$'s strategy is affected by the payoff of all agents in his Moore neighborhood (inside the dashed line); for example agent $y$, whose payoff depends of his own Moore neighborhood (shaded region). Therefore, the strategy of the agent $x$ depends on all $25$ agents in his second Moore neighborhood (inside the bold line).}
\label{Moore2Niegh}
\end{figure}

The evolutionary mechanism used is a pairwise imitation process where each agent copies the strategy of the one with maximum payoff among its neighbors. This game can be viewed as a cellular automaton, but one with $2^{25}$ rules, that's because there are $25$ neighbors (with Moore neighborhood) that affect an agent strategy at the next iteration, see Fig~\ref{Moore2Niegh}.

Even though the ``ever-changing sequences of spatial patterns" that May and Nowak saw in \cite{SpatialChaos} for a parameter regime of the game have an obvious chaotic nature, no method has been provided to evaluate or quantify the stability of a less obvious case. 

Lyapunov's exponents have been used broadly in the context of nonlinear dynamics and chaotic systems to quantify chaos. Some efforts to bring this chaos indicator to the context of cellular automata have been done in \cite{MismatchCAPrimero,MismatchCAResumen}. Indeed our PD game is basically a cellular automata as we have previously mentioned. We could use this "Lyapunov's exponent for cellular automata", in fact, we tried, but no reasonable results were obtained. We think that may be because the exponent developed in these works does not assess chaotic behaviour, but a different measure. 

To assess correctly the chaotic behaviour of the spatial patterns arising in the PD game we have used the Hamming distance. This is a measure of the difference between two ordered sets of equal length. It is defined as the number of positions at which the corresponding symbols are different in the two ordered sets. In the context of social games, Hamming distance has been used with various purposes. Authors of \cite{MinorityGame} use Hamming distance to measure the separation between strategies that are binary strings. In a similar manner, in \cite{TagBasedPGG} Hamming distance is used in order to measure the separation between a tag which agents use to discriminate other agents and either cooperate if their tag is close by or defect if its too different. 

Moreover, the Hamming distance is used to asses chaotic behaviour in \cite{HammingChaos1,HammingChaos2}, where they calculate the distance between two initially close configurations of generalized rock-paper-scissors models that describe stochastic network simulations of the May and Leonard type. They find that Hamming distance converges to a certain value, but, if a parameter representing the mobility is above a critical point, after some time, the distance oscillates increasing its amplitude of oscillation until it goes to zero or to its maximum value, the size of the population. This means that changing only one specie, or strategy, can drastically change the outcome. From a final state where one specie completely dominates the population to another final state where it is another specie that wins. This shows that the system is susceptible to initial perturbations and behaves chaotically. Our study imitates this algorithm using the Hamming distance between two initially close configurations.

In \cite{HammingChaos2} and in our context, the Hamming distance is the number of agents that have a different strategy in each configuration, which we call mismatches. We have replicated the game in \cite{SpatialChaos} and have measured the Hamming distance versus time from two relaxed configurations only differing by one agent. For most parameter values, the Hamming distance  grows to a very small constant, a periodic value, or is zero. But, for parameter values within the spatial-chaos regime discovered by the authors, the distance grows rapidly towards the Hamming distance that two random configurations would have. This rapid increase of the Hamming distance is proof of the sensitivity to the initial conditions, which affects the spatial structure of the final state. However, the mean final ratio between cooperators and defectors does not change from the two configurations, so the system only presents spatial chaos.

We have also studied a \textit{public goods game} (PGG), which presents a broader regime of parameters that present spatial chaos. This game is a prisoner's dilemma under some parameter values, but one that is played in groups of more than just two agents. The agents are either cooperators or defectors. Cooperators enrich the public goods investing a cost of $1$ unit which is reduced from their payoff. Then, for each cooperator in the group, each agent receives a quantity $r/G$, being $r$ a parameter called enhancement or multiplication factor and $G$ the number of agents in the group, which is added to their payoff either if they are cooperators or defectors. The dilemma is, therefore, that defectors leech off cooperators, reducing the global payoff. Nonetheless, cooperation can also be promoted through reputation and network reciprocity, as in the prisoner's dilemma. Additionally, cooperation can be sustained through punishment \cite{Punishment1,Punish2} and reward \cite{Reward}. We applied the imitation rule to the game but this time only one random agent is selected to change its strategy to the one of a random neighbor if its strategy is lower than the neighbor's. By setting all agents in a square lattice and making them play games with its immediate neighbors, a spatial structure appears. These structures are not static, but evolve in time, and so we want to measure how chaotic this pattern evolution is.  Authors of \cite{PatternECOPGG} have studied the ecological PGG and have found chaotic dynamics similar to the patterns of \cite{SpatialChaos} and they have verified the irregular dynamics qualified as spatial chaos. 

For both cases, we have computed the Hamming distance over time, and normalized to the statistical Hamming distance of two random configurations of cooperators and defectors with the proportions matching the studied system at each time. Our main finding indicates that the divergence of two initially close solutions, measured using the Hamming distance, serves as a reliable indicator of chaotic dynamics in the deterministic game under study. Furthermore, in the case of the game exhibiting a degree of stochasticity, the distance increases more rapidly in regimes displaying spatial-chaotic behavior.

The organization of this manuscript is as follows. On Section~\ref{PrisonersDilemaSection}, we replicate the results of Novak and May \cite{SpatialChaos} and calculate the normalized Hamming distance fitting it with the Weibull ''stretched exponential" function. On Section~\ref{PGGSection}, we explain the model used for the public goods game and also fit the normalized Hamming Distance. Finally, we present the main conclusions at the end.

\section{Prisoner's dilemma}
\label{PrisonersDilemaSection}

\begin{figure}
    \centering
\begin{tikzpicture}

\draw[very thick] (-3.2,0) -- (3.2,0);
\draw[very thick] (0,-3.2) -- (0,3.2); 

\draw[very thick] (-0.2,3) -- (0,3.2);
\draw[very thick] (0.2,3) -- (0,3.2);

\draw[very thick] (3,-0.2) -- (3.2,0);
\draw[very thick] (3,0.2) -- (3.2,0);

\draw (-0.5,1.6) node {\Large{$D_g$}};
\draw (1.6,-0.4) node {\Large{$D_r$}};

\draw (-1.6,1.6) node {\Large{\textbf{CH}}};
\draw (1.6,1.6) node {\Large{\textbf{PD}}};
\draw (1.6,-1.6) node {\Large{\textbf{SH}}};
\draw (-1.6,-1.6) node {\Large{\textbf{H}}};

\end{tikzpicture}
    \caption{Diagram representing the four types of pairwise social games according to dilemma strength. CH stands for the Chicken game, PD for the Prisoner's Dilemma, SH stands for the Stag-hunt game and H stands for Harmony game. }
    \label{DiagDilemma}
\end{figure}
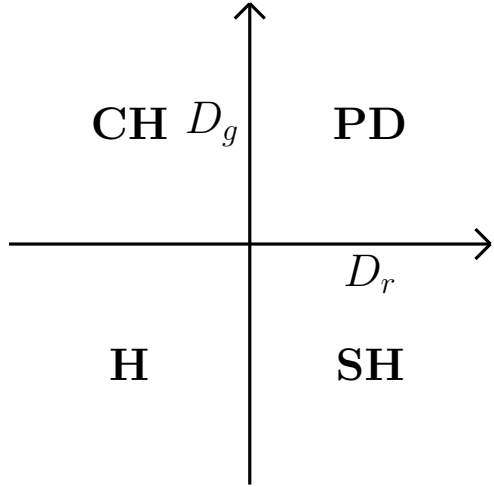

\begin{figure}
	\centering
	\includegraphics[width=1\linewidth]{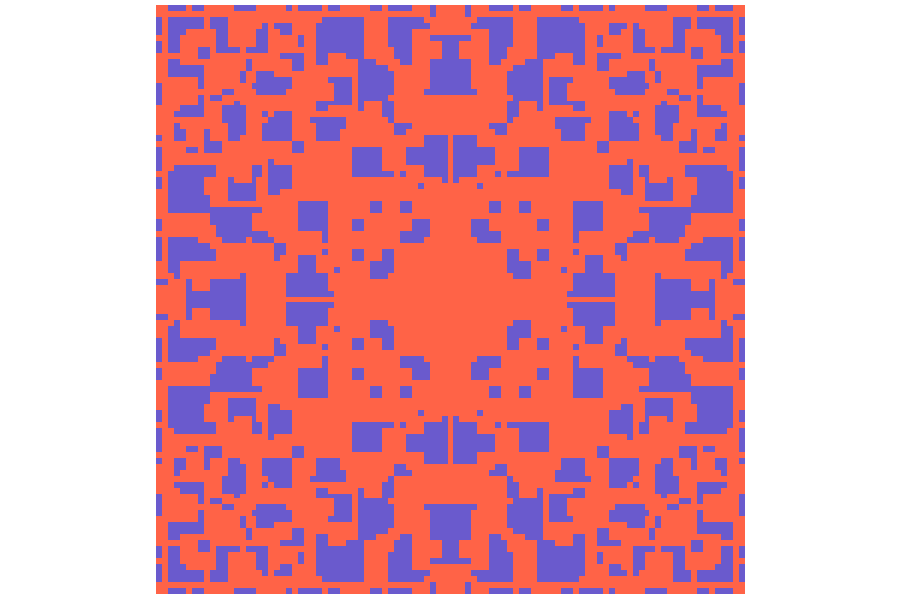}
	\caption{Color map of the prisoner's dilemma with a Moore neighborhood, $R=1$, $P=S=0$, and $1.8<T<2$. We have represented cooperators in blue and defectors in red for a grid size of $L=99$ and fixed boundary conditions after relaxation of the prisoner's dilemma. The box counting dimension of the boundary between the region of cooperators and defectors is $2$. }
	\label{PDcolormapA}
\end{figure}

\begin{figure}
	\centering
	\includegraphics[width=1\linewidth]{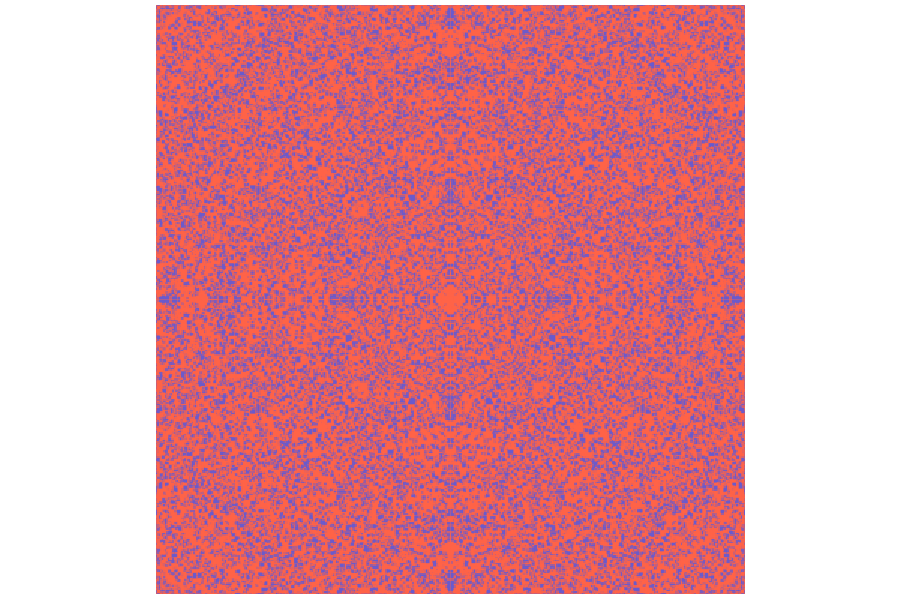}
	\caption{Color map of the prisoner's dilemma with a Moore neighborhood, $R=1$, $P=S=0$, and $1.8<T<2$.  We have represented cooperators in blue and defectors in red for a grid size of $L=999$ and fixed boundary conditions after relaxation of the prisoner's dilemma. Box counting dimension of the frontier between defectors and cooperators is $2$>. There is not seen clusters of the same strategy at different scales because all clusters have similar size. In other words, the boundary is not fractal.}
	\label{PDcolormapB}
\end{figure}

The game in~\cite{SpatialChaos} is the spatial prisoner's dilemma (PD). It is an agent-based model where the authors put agents in the nodes of a square lattice and each agent played a game with each of its Moore neighbors. The total payoff is the sum of the payoff gained in each game which was $R=1$ if both players cooperate, $P=0$ if both players defect, and if an agent defects and another cooperates, the first gains $T=b>1$ and the second $S=0$. Even though the prisoner's dilemma condition $T>R>P>S$ does not hold strictly, the conclusions are the same for $P=\epsilon\to0$ positive, which satisfies the condition. 
The concept of dilemma strength \cite{DilemmaStrength1,DilemmaStrength2,DilemmaStrength3} is a key classification of pairwise social games. This is defined by the risk-averting dilemma strength $D_r=P-S$ and the gamble-intending dilemma strength $D_g=T-R$. The sign of these two values distinguishes the social games in four types of games as seen on the diagram of Fig.~\ref{DiagDilemma}. 

The settings of the game on \cite{SpatialChaos} are that of a boundary game, since $D_r=0$. This type of game is not a well representation of the general PD since it has none of Stag-Hunt type dilemma, but has the advantage of having a single parameter representing dilemma extent. A more suited dilemma setting that also has the advantage of having a single parameter representing dilemma extent is the Donor \& Recipient game, presumed in the community of theoretical biologist as the standard template representing PD \cite{DonorRecipient}. Besides this, we have chosen to adopt the boundary game from Nowak and May to replicate the same results at the same parameter values, and thus, see if our tools correctly measure spatial chaos.

At each iteration of the game, each agent copies the strategy of its Moore neighbor that has the highest payoff, including themselves. The process happens simultaneously for all agents. Therefore, the game is deterministic, and its rules are symmetrical. Granting that the initial conditions have symmetry, e.g., all agents are cooperators except one defector at the center, the spatial distribution of cooperators and defectors will be symmetrical. Nonetheless, the generated patterns are not fractals as the authors claim. In Figs.~\ref{PDcolormapA} and \ref{PDcolormapB} we plot cooperators in blue and defectors in red for different population sizes $N=L^2$, being $L$ the grid size. Observing Fig.~\ref{PDcolormapA} that corresponds to the case $L=99$, one could think that the structure is fractal as the authors claim in~\cite{SpatialChaos}. However, the pattern is not fractal as we can see on Fig.~\ref{PDcolormapB} for $L=999$, instead we can observe near homogeneous size clusters. Furthermore, we computed the box counting dimension of the boundary between cooperators and defectors and it is $2$ for both figures.

\begin{figure}
	\centering
	\includegraphics[width=1\linewidth]{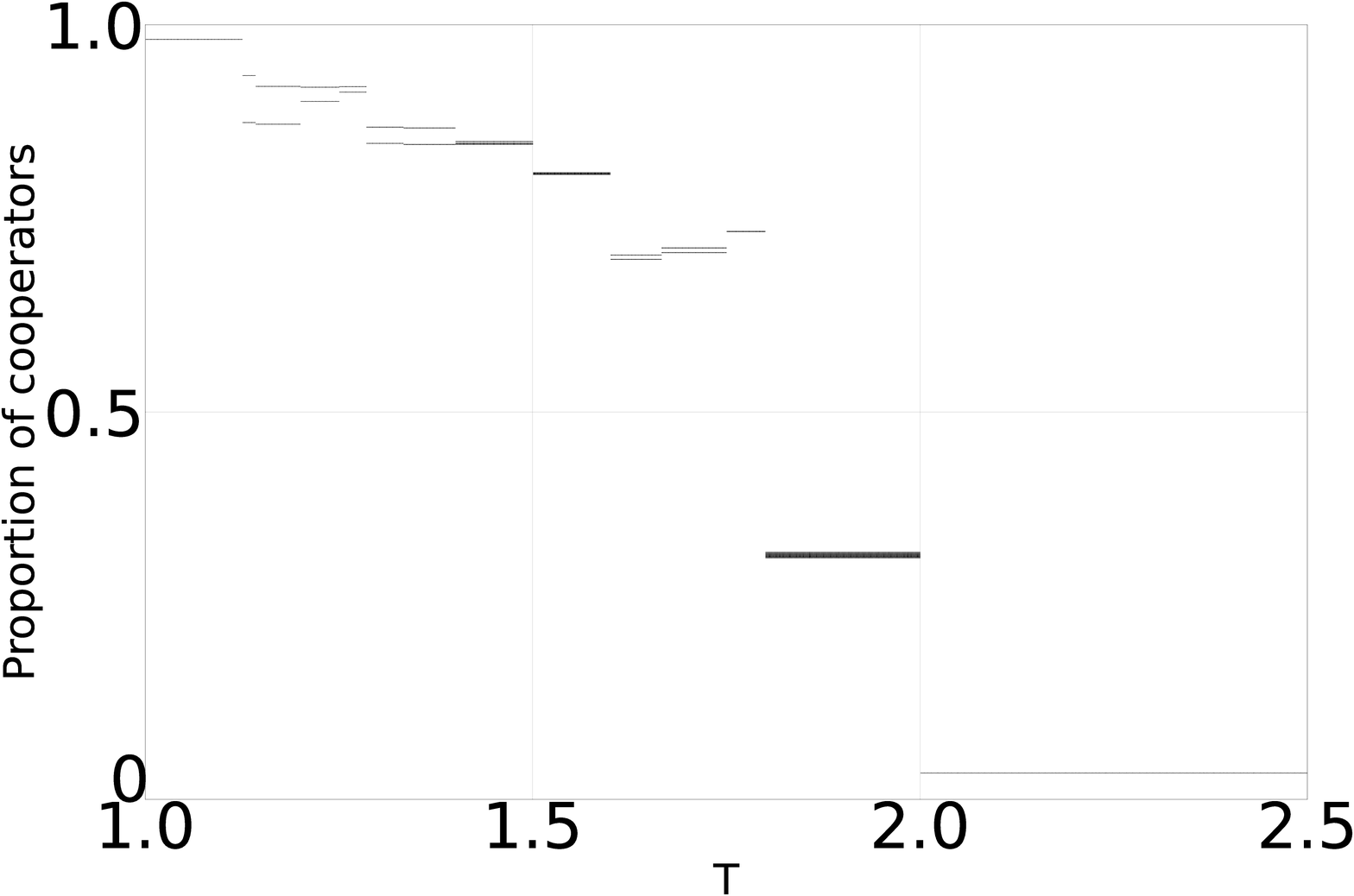}
	\caption{Proportion of cooperators for the last $50$ iterations after relaxation of a deterministic prisoner's dilemma game with Moore neighborhood and a population size of $999\times999$ in function of the temptation payoff parameter $T$. We can distinguish 14 regions corresponding with shifts at the values of the temptation payoff that induce a change in the game. At $1.8<T\leq2$ the dynamics is chaotic. As we can see, the line is bold, i.e., there is a large number of possible states with different proportion of cooperators, whereas for the other regions only a few periodic states are present.}
	\label{PDbifurcation}
\end{figure}

In Fig.~\ref{PDbifurcation} we show the proportion of cooperators for the last $50$ generations after relaxation of a deterministic PD game with a Moore neighborhood and $L=999$ in function of the temptation payoff $T$. We can see that the $1.8<T<=2.0$ regime presents a different behavior, than the rest of the parameter scan, a chaotic one. There is a sudden transition to chaos due to the discrete nature of the temptation payoff $T$, i.e., although $T$ is a real number, the system is only altered at some thresholds by the nature of the game rules, and not continuously. Assuming that two neighboring agents, denoted as $i$ and $j$, employ different strategies, if agent $j$, for instance, alters its strategy to align with that of agent $i$, it is inferred that the payoff for agent $i$, denoted as $\Pi_i$, surpasses that of any neighbors of agent $j$. Assume that the agent $i$ has $x$ neighbouring cooperators and that the neighbor of agent $j$ with maximum payoff besides agent $i$, which we will call agent $k$ with payoff $\Pi_{k}$, has $y$ neighbouring cooperators. The thresholds are given by the possible values of $y/x$. This is because $\Pi_{k}=y$ if agent $k$ is a cooperator or $\Pi_{k}=yT$ if it is a defector. Likewise $\Pi_i=xT$ if agent $x$ is a defector or $\Pi_{i}=x$ if it is a cooperator, so $\Pi_i>\Pi_{k}$ is held by $T>y/x$ or $T<x/y$. The value of $T$ has to be greater than one for the game to be a prisoner's dilemma, and $x,y\leq9$ since there are only $9$ agents that play each game. Therefore there are $27$ thresholds, but in fact, in Fig.~\ref{PDbifurcation} defectors completely dominate the game for values beyond $T>2$, so we would only see $15$ different regimes. We only see $14$ regimes in Fig.~\ref{PDbifurcation} since there is no change in the frequency between threshold $7/6$ and $6/5$ for unknown reasons.

\subsection{Hamming distance}

The Hamming distance between two configurations $s$ and $s'$ of cooperators and defectors is 
\begin{equation}
    H(t)=\sum_i^N |s_i-s'_i|,
\end{equation}
where $s_i$ and $s'_i$ is the strategy of agent $i$ in the configuration $s$ or $s'$ which can be $0$, a defector, or $1$, a cooperator.

We calculate the final Hamming distance that separates two configurations that begun at the same initial conditions, where there is a random 50\%-50\% cooperator-defector setting, with only one random-agent's strategy changed from one configuration to the other. We normalize this distance to the statistical Hamming distance, which is the Hamming distance that separates two random configurations with proportion of cooperators $p_C(t)$ and defectors $p_D(t)$, which can be calculated as 
\begin{equation}
H_{stat}(t)=p_C(t)p_D(t)+p_D(t)p_C(t)=2p_C(t)(1-p_C(t)).
\end{equation}

The normalized Hamming distance in our context is expressed as $H'(t) = H(t) / H_{stat}(t)$. It is essential to distinguish this normalization method from the conventional approach applied to the Hamming distance, which typically involves normalization based on the size of the set.

\begin{figure}
	\centering
	\includegraphics[width=1\linewidth]{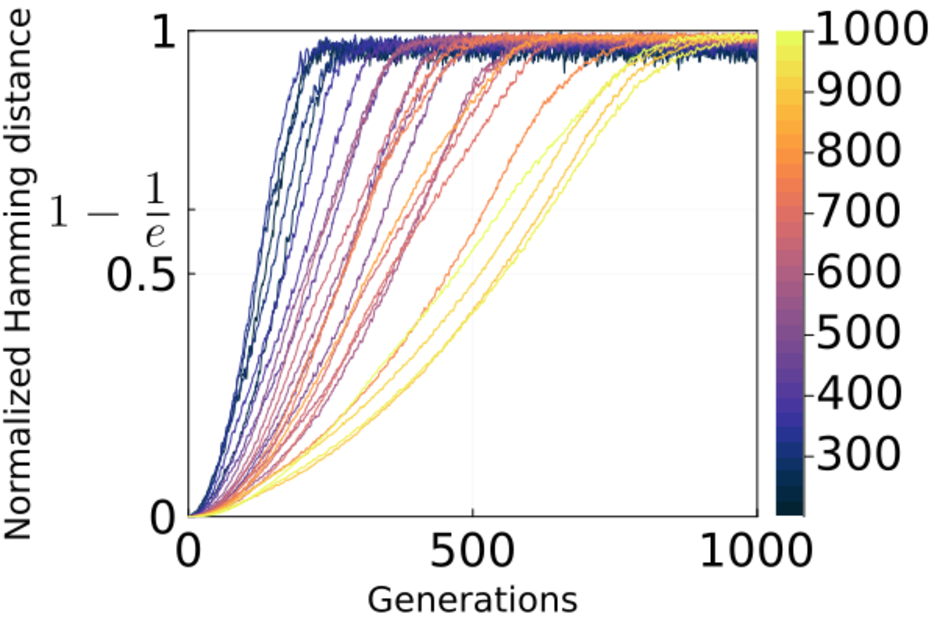}
	\caption{ Normalized Hamming distance of the two solutions versus time (number of generations). Multiple curves are shown with different colors, representing the different grid size $L$ values. The curves grow in a sigmoid-like curve towards one. They are normalized to the statistical Hamming distance which depends on L. The larger $L$ is, the longer it takes for the normalized Hamming distance to reach $1$.}
	\label{PD_NormalHammingVSt}
\end{figure}

\begin{figure}
	\centering
	\includegraphics[width=1\linewidth]{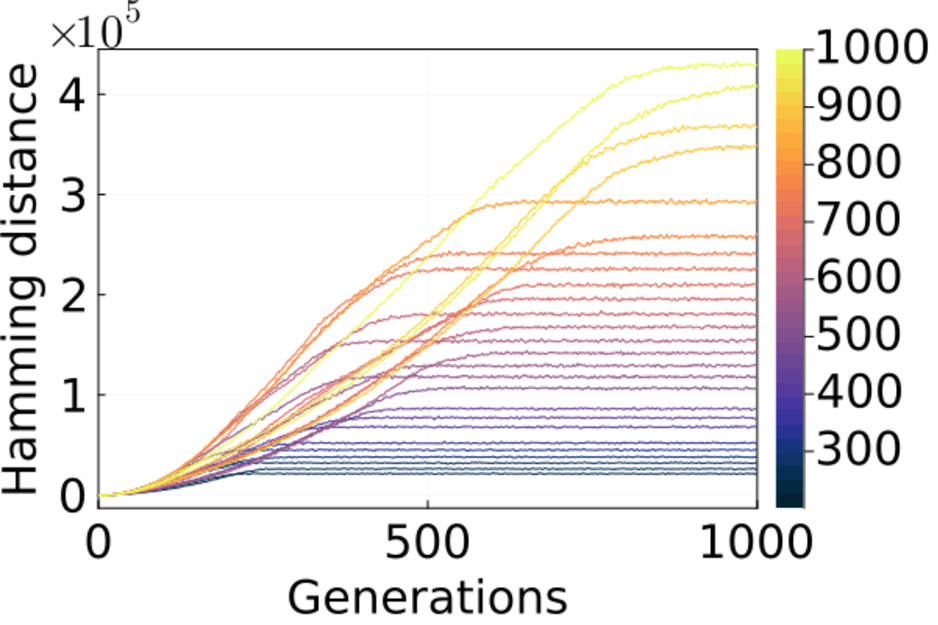}
	\caption{Hamming distance of the the two solutions versus time (number of generations). Different colors represent different values of the grid size $L$. The larger the grid size, the longer it takes for the Hamming distance to reach $H_{stat}$.}
	\label{PDHammingVst}
\end{figure}

We have seen that the final normalized distance is zero (or close to zero) for all values except for the parameter regime specified by Novak and May in \cite{SpatialChaos}, where it is close to one. This indicates that this parameter regime presents spatial chaos, whereas the others do not. In Fig.~\ref{PD_NormalHammingVSt} we plot $H'(t)$ and it grows as a sigmoid-like curve. For larger populations sizes it takes more time to reach the maximum value of the distance, but it gets closer to $H_{stat}$ for large times. This happens because $H_{stat}$ grows as the population size increases, but the growing rate of the Hamming distance is not altered by the population size, as seen on  Fig.~\ref{PDHammingVst}, where all curves have a similar growing rate. Discrepancies in the growing rate are due to the random initial conditions and the choice of the first mismatch, i.e., the first agent that is different in both configurations, which is also randomly chosen.

We have fitted these curves to a sigmoid function. We have chosen the Weibull ``stretched exponential" function \cite{Weibull}, of the form
\begin{equation}
    F(t;k,a)=1-e^{-(t/a)^k},
    \label{WeibullDistr}
\end{equation}
where $a$ sets the timescale, and $k$ indicates how abruptly the curve grows. We choose this fit since it is sigmoid-like, it starts at the origin and has only two relevant parameters.

\begin{figure}
	\centering
	\includegraphics[width=1\linewidth]{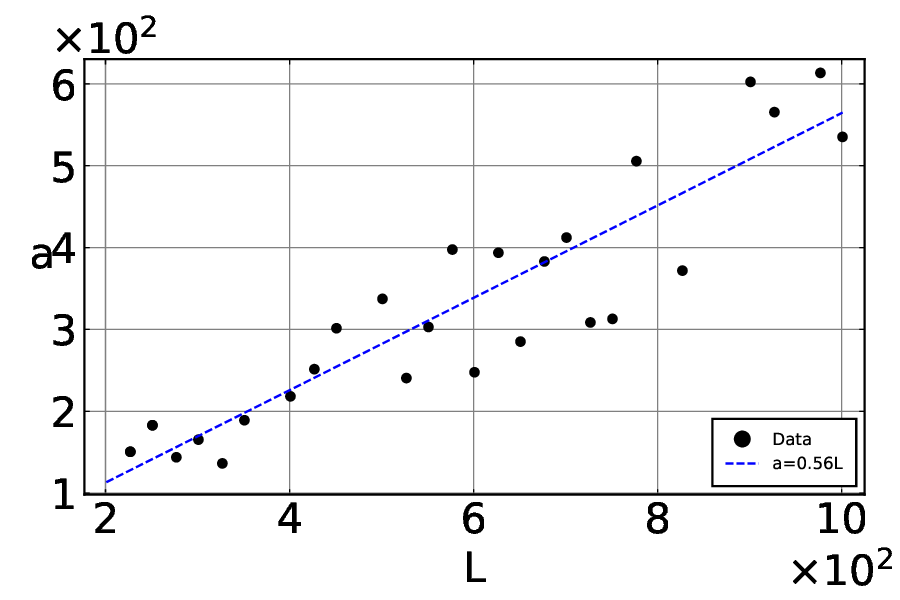}
	\caption{Parameter $a$ from the Weibull ``stretched exponential" function $F(t;k,a)=1-e^{-(t/a)^k}$ fitted to the normalized Hamming distance of the solutions for the prisoner's dilemma versus grid size $L$. It shows a linear regression (dashed blue line) where $a$ grows proportionally to $L$. Error bars, derived from the fitting, are smaller than marker size.}
	\label{PDaVsL}
\end{figure}

\begin{figure}
	\centering
	\includegraphics[width=1\linewidth]{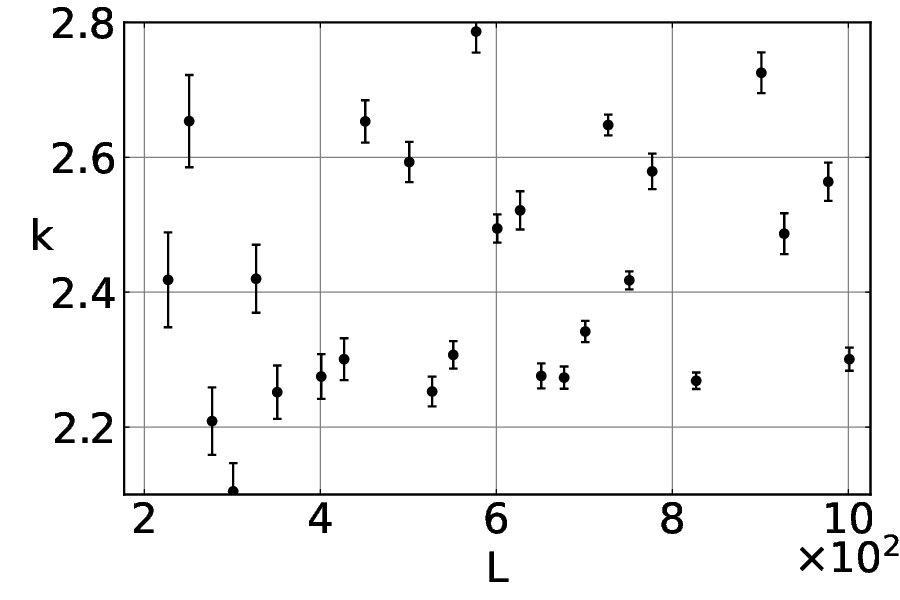}
	\caption{Parameter $k$ from the Weibull ``stretched exponential" function $F(t;k,a)=1-e^{-(t/a)^k}$ fitted to the normalized Hamming distance of the solutions for the prisoner's dilemma versus grid size $L$. All points have similar values of $k \approx 2.4\pm0.4$. Error bars are derived from the fitting.}
	\label{PDkVsL}
\end{figure}

Figure~\ref{PDaVsL} shows that $a$ grows linearly with $L$ whith a slope of $0.56 \pm 0.05$ and the intercept is $0 \pm 30$. It is reasonable that $a$ grows linearly with $L$, since the mismatches can propagate at a velocity limited by the range of action of an agent, i.e., the distance at which an agent can affect another agent. We can see in Fig.~\ref{Moore2Niegh} that this distance is two agents at each direction, so the time it takes a mismatch to propagate to the whole population should be no less than $0.5L$ iterations. Since the slope is near this maximum of velocity, it shows that the propagation of mismatches is very fast and uninterrupted. As we see in Fig.~\ref{PDkVsL}, the value of $k$ is independent of  $L$, all values are around $2.45\pm0.4$.

\section{Public Goods game}
\label{PGGSection}

\subsection{Model and initial results}

\begin{figure}
\centering
\begin{tikzpicture}

\fill[gray!40!white] (2,2) rectangle (3,5);
\fill[gray!40!white] (1,3) rectangle (4,4);

\draw[step=1cm,gray,very thin] (0,0) grid (5,5);

\draw[very thick] (0,2) -- (0,3) -- (1,3) -- (1,4) -- (2,4) -- (2,5) -- (3,5) -- (3,4) -- (4,4) -- (4,3) -- (5,3) -- (5,2) -- (4,2) -- (4,1) -- (3,1) -- (3,0) -- (2,0) -- (2,1) -- (1,1) -- (1,2) -- (0,2);

\draw (2.5,2.5) node {\Large{\textbf{i}}};

\end{tikzpicture}
\caption{Von Neumann neighborhood at a distance $2$ of agent $i$. The individual $i$ plays $5$ games with agents in cross-like patterns like the one shaded.}
\label{2neighbors}
\end{figure}
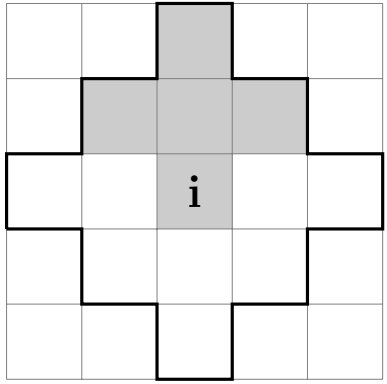

We have studied the PGG by means of an agent-based model of $N$ total agents where each one is a node on a square lattice that plays $5$ games with $G=5$ von Neumann second nearest neighbors, those agents at a Manhattan distance of two nodes, which form cross-like patterns like that of Fig.~\ref{2neighbors}. In each game $g=1,..., G$, cooperators gain a payoff $\Pi^g_C$ and defectors gain a payoff $\Pi^g_D$ that depends on the number of cooperators $N^g_C$ among their neighbors

\begin{equation}
\begin{split}
&\Pi_C^g=\frac{r}{G}N_C^g-1 \\
&\Pi_D^g=\frac{r}{G}N_C^g.
\end{split}    
\end{equation}

The accumulated payoff $\Pi=\sum_g^G \Pi^g$ of each agent, calculated as the sum of the payoff gained in each game played, determines how likely each agent survives and reproduces. The evolutionary mechanism we have adopted is a pairwise imitation process in which, through Monte Carlo simulations, an agent is randomly chosen to adopt the strategy of a random nearest neighbor if its payoff is lower than the neighbor's. Cooperation is chosen if there is a tie in payoffs. 

Once $N$ iterations has passed, all agents have had the possibility on average to adopt a neighbor's strategy. Hence our Monte Carlo Step (MCS), which accounts for one generation, will be $N$ iterations.

We have started every simulation with each agent strategy chosen at random with the same probability of being cooperators or defectors.

\begin{figure}
	\centering
	\includegraphics[width=1\linewidth]{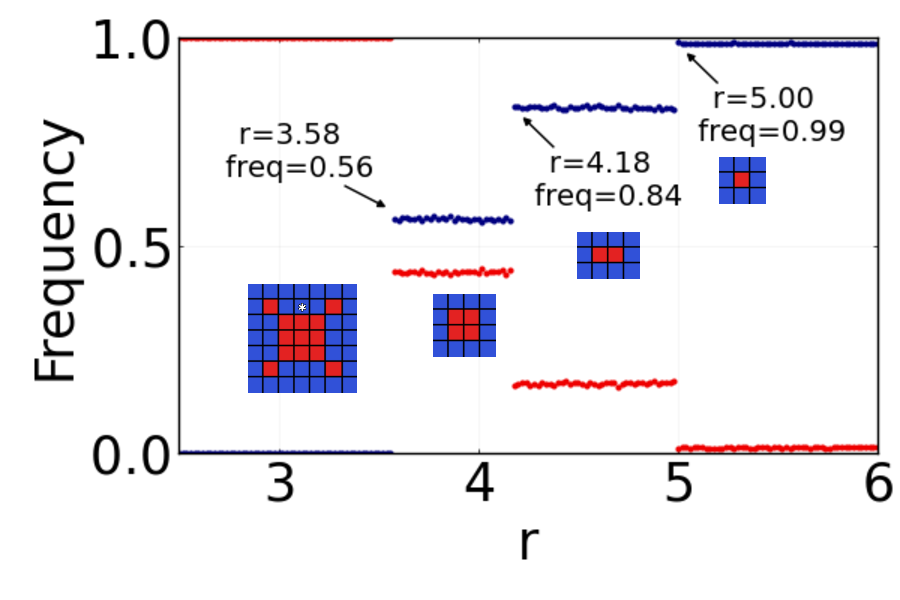}
	\caption{Proportion of cooperators (in blue) and defectors (in red) after running a simulation of the public goods game starting with a random strategy in each of the $N=90000$ agents for a relaxation time of $2000$ generations. We can see a series of steps matching the values of $r$ that, for the given configurations, the surrounding cooperators have greater payoff than the surrounded defectors. For the first configuration this holds strictly for the white-marked cooperator and the defector below it at the first shift at $r=25/7\approx3.57$.}
	\label{freqK0}
\end{figure}

After running the simulation through a relaxation time of $2000$ generations, we present the proportion of cooperators and defectors versus the multiplication factor $r$ in Fig.~\ref{freqK0}. In this figure, we see a series of shifts discontinuously increasing cooperation proportion, and horizontal steps where the proportion remains constant. We are able to predict the $r$ values for which the proportion of cooperators shifts by calculating the payoffs of the surrounding cooperators and defectors in the shown configurations.

We can get the value of the $r$ value of the first shift calculating the payoff of the white-marked cooperator and the defector below it. The payoff of this cooperator is
\begin{equation}
    \Pi_C=\frac{1}{5}(5r+4r+2\times3r+r)-5=\frac{16}{5}r-5,
\end{equation}

The payoff of the defector is
\begin{equation}
    \Pi_D=\frac{1}{5}(4r+2\times2r+r+0)=\frac{9}{5}r.
\end{equation}

The shifts in Fig.~\ref{freqK0} occur when there is a change in the behavior of the system. The shifts are met when $\Pi_C=\Pi_D$, since cooperation gets more profitable. For the first configuration this shift occurs at $r\geq25/7\approx3.57$. The rest of shifts occur at $r\geq25/6=4.1\hat6$, $r\geq25/5=5$ and $r\geq25/3=8.\hat3$ all well in accordance with the figure. The last shift has not been plotted on the figure because is barely visible, but at this point cooperators completely dominate the game.

In the PGG cooperation is a Nash equilibrium for $r\geq5$. Nash equilibrium occurs when no agent can increase its payoff when changing strategy if the others keep theirs unchanged. Nonetheless since the payoff of each agent is affected by the strategies of its neighbours, a change of one's strategy can lower the payoff of its neighbours more than what it lowers his own payoff, making Nash equilibrium not being a sufficient condition to eliminate all defectors. For a cooperative Nash equilibrium with $r\geq5$ even a defector surrounded by cooperators would have greater payoff if it were a cooperator, but since its surrounding cooperators have lesser payoff, it cannot adopt the cooperation strategy given our evolutionary model and therefore frequency is lower than 1, see Fig.\ref{freqK0}. This holds until $r=8.\hat3$ as the theoretical study predicts, when cooperators have more payoff than the isolated defector. We have not plotted this shift because it can hardly be appreciated at the current scale and the computational times for all defectors to be gone are large.

\subsection{Hamming distance}

We wanted to know how fast two solutions for the public goods game that differed only on one agent's strategy at the beginning would separate calculating their Hamming distance, so we could get an idea of when the system presents spatial chaos. The election of the agent subjected to adopt a new strategy is chosen at random by the Monte Carlo simulation. This stochasticity, could be manifest in a growth of the Hamming distance. We try to minimize this effect by setting all random interactions to be a set of predetermined values, the same for both configurations but different at each iteration. 

\begin{figure}
	\centering
	\includegraphics[width=1\linewidth]{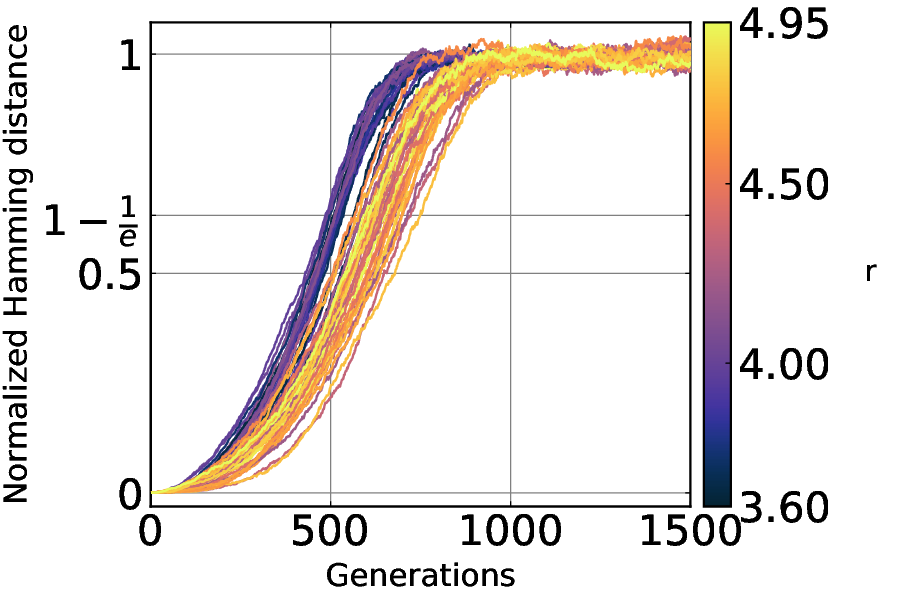}
	\caption{Normalized Hamming distance between the solutions for the public goods game versus time (number of generations). Multiple curves are shown with different colors, representing the different $r$ values. The curves grow in a sigmoid-like curve towards one. They are normalized to the statistical Hamming distance, which depends on $r$, so curves ranging from  $25/7\leq r<25/6$ are normalized to a different value than those at $25/6\leq r<5$. There is a distinction between the two regimes, the curves of the first one (blue ones) reach its midpoint after than than the second regime (orange ones).}
	\label{PGGNormalizedHammvst}
\end{figure}

\begin{figure}
	\centering
	\includegraphics[width=1\linewidth]{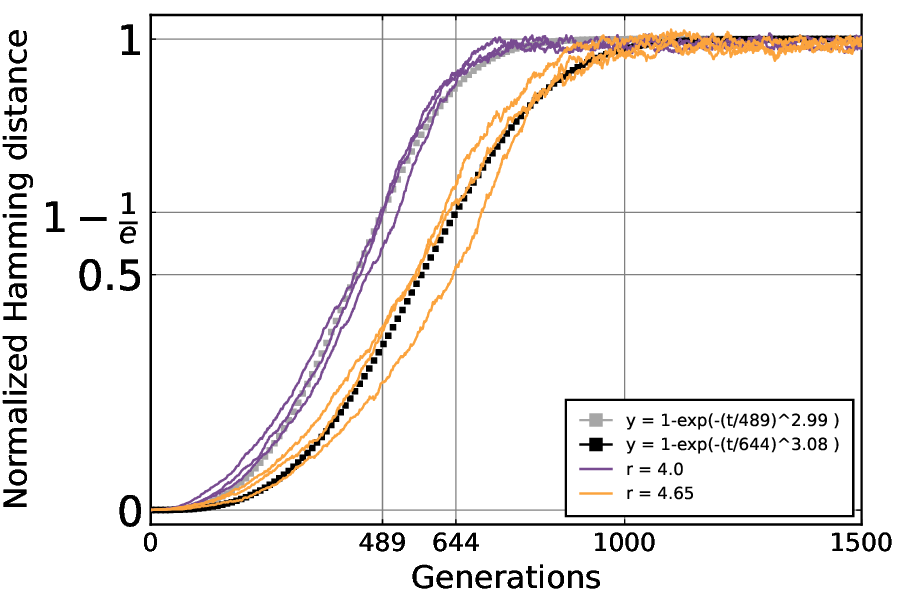}
	\caption{Normalized Hamming distance between the solutions for the public goods game versus time (number of generations). Different colors represent the different $r$ values. The curves grow towards one in a sigmoid-like form. They are normalized to the statistical Hamming distance, which depends on $r$, so the two curves are normalized to a different value. They are fitted with the Weibull ``stretched exponential" function $F(t;k,a)=1-e^{-(t/a)^k}$. The $r=4$ curves (blue and left ones), representing the regime $25/7\leq r<25/6$ has a lower value for the $a$ parameter than the curves with $r=4.65$ (orange and right ones), meaning that for this parameter, the system is more sensitive.}
	\label{PGGNormalizedHammvstFitted}
\end{figure}

\begin{figure}
	\centering
	\includegraphics[width=1\linewidth]{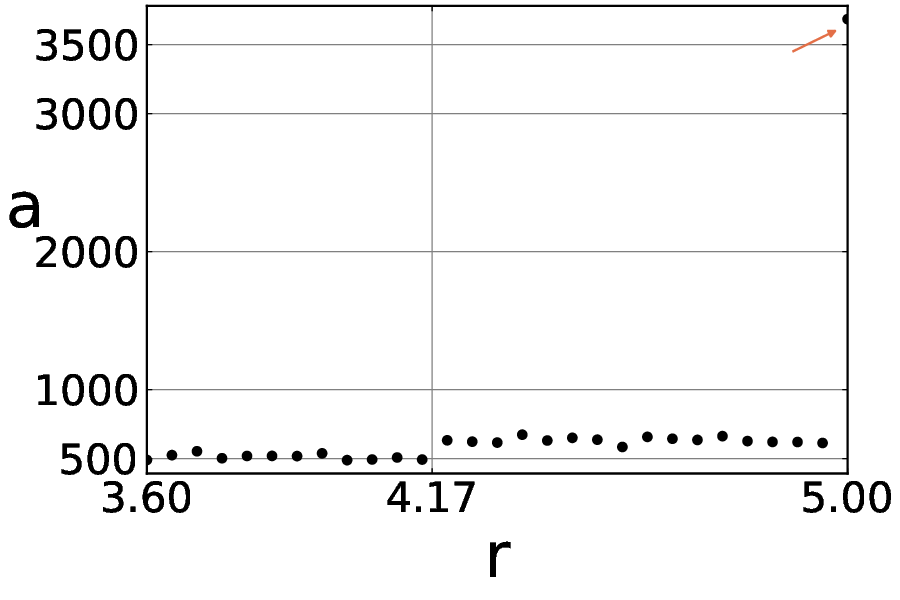}
	\caption{Value of the fitted parameter $a$ in the Weibull ``stretched exponential" function $F(t;k,a)=1-e^{-(t/a)^k}$ fitted to the normalized Hamming distance of two solutions for the public goods game for different $r$ values. Notice the rightmost point at $r=5$ and $a\simeq3600$. We can see the shift between the 3 regimes $25/7\leq r<25/6\approx4.17$, $25/6\leq r<5$ and $r\geq5$. Error bars, derived from the fitting, are smaller than marker size. }
	\label{Fittedavsr}
\end{figure}

\begin{figure}
	\centering
	\includegraphics[width=1\linewidth]{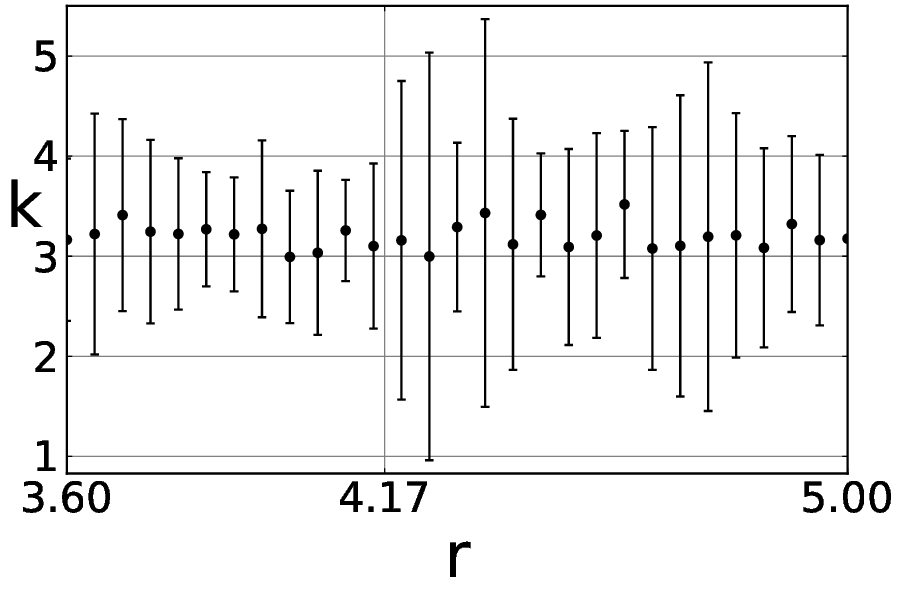}
	\caption{Value of the fitted parameter $k$ in the Weibull ``stretched exponential" function $F(t;k,a)=1-e^{-(t/a)^k}$ fitted to the normalized Hamming distance of two solutions for the public goods game with different $r$ values. All points exhibit comparable values of $k$, approximately around $k \approx 3.2$, and are indistinguishable across various regimes. Error bars are derived from the fitting.}
	\label{Fittedkvsr}
\end{figure}

We fit the normalized Hamming distance with the Weibull ''stretched exponential" function as in the previous section. In Fig.~\ref{PGGNormalizedHammvst}, we plot the normalized Hamming distance for different $r$ values. We see two slightly separated regimes. Those in $25/7<r<25/6$, which have lesser $a$ values and those in $25/6<r<5$, with larger $a$ as fitted in Fig.~\ref{PGGNormalizedHammvstFitted}. 
This is well observed in Fig.~\ref{Fittedavsr}. However, the value of $k$ seems to be independent of $r$ as can be seen in Fig.~\ref{Fittedkvsr}. All curves have similar $k$ values of $3\pm1$. In \cite{HammingChaos2} they refer to the time it takes for the Hamming distance to reach $1/4$ of the maximum, which is similar as the value of $a$, and compared it to the Lyapunov time, but never developed further investigation.

\begin{figure}
	\centering
	\includegraphics[width=1\linewidth]{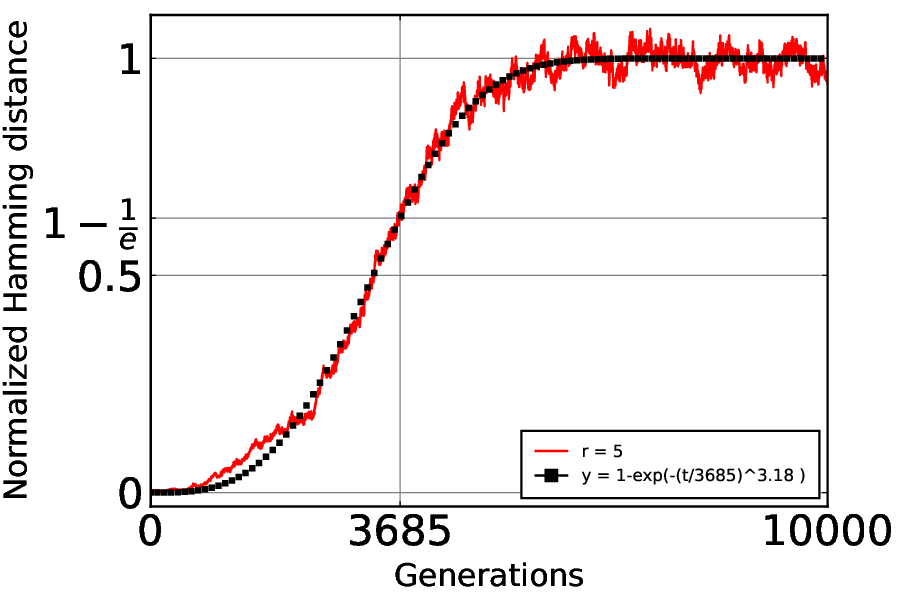}
	\caption{Normalized Hamming distance between the solutions for the public goods game versus time (number of generations). The curve grows in a sigmoid-like curve towards one. It is normalized to the statistical Hamming distance and it is fitted with the Weibull ``stretched exponential" function $F(t;k,a)=1-e^{-(t/a)^k}$. This curve for $r=5$ is plotted separated from the previous ones because it spans farther into the x axis since the value of $a$ is far greater, meaning the system is less unstable for this parameter value.}
	\label{PGGNormalizedHammvst_r5}
\end{figure}

In Fig.~\ref{PGGNormalizedHammvst_r5}, we see that for $r=5$, $a$ is broadly larger than at the other regimes. However, the Hamming distance still grows towards $H_{stat}$, even though we did not expect spatial chaos for this parameter value because of the sparse density of defectors. This could be a sign that the Hamming distance also tracks the randomness of the Monte Carlo simulations. Since we set all random interactions at the beginning, the agent that adopts a new strategy is the same for the two solutions, but since one solution initially differs from the other, the outcomes become different and diverse, diverging more at each iteration. However, we get a much larger $a$ value, meaning that it takes much more time for the Hamming distance to reach its maximum so we perceive this as a sign that the system is less unstable. This gives us the idea that the system with this parameter value is less spatially chaotic, if chaotic at all.

\section{Conclusions}
\label{ConclusionsSection}

We have studied the public goods game to see how two relaxed almost identical configurations diverged in its spatial configuration through the means of the Hamming distance in order to categorize if there was spatial chaos. Under the study of a well-known spatial chaos example, the game of Nowak and May \cite{SpatialChaos}, we have seen that the Hamming distance grows as a sigmoid-like curve towards the statistical Hamming distance that two random configurations would have for those cases where spatial chaos was observed, and the distance is zero or close to zero for the cases where the system falls into a fixed point or periodic solutions. The settings of the prisoner's dilemma game were that of a boundary game where $D_r=0$ which is not the general case for PD, but we used it nonetheless since we wanted to replicate the same results as in~\cite{SpatialChaos}. We have no reasons to think the main conclusions will not hold for a more general case.

For the public goods game, the Hamming distance grows towards the statistical value for all the studied parameter values, even for those we did not expect the system to be spatially chaotic, that is, with $r>5$. This behavior may be explained due to the randomness of the Monte Carlo simulation, making the instability test less precise even though we have tried to minimize this by setting the same random values in both configurations. Moreover, the algorithm is not completely useless since the distance grows much slower for those cases. Therefore our tool gives a broad idea of how unstable are the spatial configurations, even though is not completely efficient in discerning a spatially chaotic behavior in a model with some sort of stochasticity, therefore further research should be done.

Our study does not replicate the evolution of the Hamming distance equally as in the case of \cite{HammingChaos2}, where initial perturbation changes which species would win changing drastically the frequency of species value. In contrary, our case keeps the same values of strategy frequencies on average. This means that the models studied does not behave chaotically concerning frequencies since they converge to an equilibrium, although they do present spatial chaos, and our studies have brought a way to discern it.

While our examination has been limited to the spatial prisoner’s dilemma and the spatial public goods game on a square lattice, we posit that the Hamming distance can be applied to assess instability across different lattice geometries in networks. Furthermore, we believe it can be extended to evaluate instability in other game scenarios.  

\section*{Acknowledgments}

This work has been supported by the Spanish State Research Agency (AEI) and the European Regional Development Fund (ERDF, EU) under Project No.~PID2019-105554GB-I00 (MCIN/AEI/10.13039/501100011033)

\end{document}